\documentclass[aps,10pt,a4paper,twocolumn,footinbib, superscriptaddress, floatfix]{revtex4-2}
\usepackage{graphicx}
\usepackage{amsmath}
\usepackage{amsfonts}
\usepackage{amssymb}
\usepackage{url}
\usepackage{hyperref}
\hypersetup{colorlinks=true, citecolor=blue, urlcolor=blue, linkcolor=blue}
\usepackage{amstext}
\usepackage[separate-uncertainty=true]{siunitx}
\DeclareSIUnit\sq{\ensuremath\Box}
\DeclareSIUnit{\sqrthz}{\ensuremath{\sqrt{\mathrm{\hertz}}}}

\usepackage{booktabs}
\usepackage{miller}
\usepackage[version=4]{mhchem}

\newcommand{\Lks}{L_{k, \square}}
\newcommand{\zzpf}{\ensuremath{ z_{\mathrm{zpf}} }}
\newcommand{\meff}{\ensuremath{ m_{\mathrm{eff}} }}

\newcommand{\kB}{\ensuremath{ k_{\mathrm{B}} }}
\newcommand{\Qtot}{\ensuremath{ Q_{\mathrm{tot}} }}
\newcommand{\Qint}{\ensuremath{ Q_{\mathrm{int}} }}
\newcommand{\Qext}{\ensuremath{ Q_{\mathrm{ext}} }}

\date{February 15, 2024}

\begin{document}

\title{Design, fabrication and characterization of kinetic-inductive force sensors for scanning probe applications}

\author{August K. Roos}
\author{Ermes Scarano}
\author{Elisabet K. Arvidsson}
\author{Erik Holmgren}
\author{David B. Haviland}
\email{haviland@kth.se}
\affiliation{Department of Applied Physics, KTH Royal Institute of Technology, Hannes Alfvéns väg 12, SE-114 21 Stockholm, Sweden}

\begin{abstract}
We describe a transducer for low-temperature atomic force microscopy based on electromechanical coupling due to a strain-dependent kinetic inductance of a superconducting nanowire. The force sensor is a bending triangular plate (cantilever) whose deflection is measured via a shift in the resonant frequency of a high-Q superconducting microwave resonator at \SI{4.5}{\giga\hertz}. We present design simulations including mechanical finite-element modeling of surface strain and electromagnetic simulations of meandering nanowires with large kinetic inductance. We discuss a lumped-element model of the force sensor and describe the role of an additional shunt inductance for tuning the coupling to the transmission line used to measure the microwave resonance. A detailed description of our fabrication is presented, including information about the process parameters used for each layer. We also discuss the fabrication of sharp tips on the cantilever using focused electron beam-induced deposition of platinum. Finally, we present measurements that characterize the spread of mechanical resonant frequency, the temperature dependence of the microwave resonance, and the sensor's operation as an electromechanical transducer of force.
\end{abstract}

\maketitle

\section{Introduction}\label{sec:introduction}

Cavity optomechanics~\cite{Aspelmeyer2014review} deals with the detection and manipulation of massive ``test objects'' at the fundamental limits imposed by quantum physics~\cite{Braginsky1975quantumlimit}. By detecting the motion of the test object we can sense an external force, for example gravitational waves acting on a \SI{40}{\kilo\gram} mirror in LIGO~\cite{Abbott2016ligo}, or atomic-scale tip-surface force acting on a \SI{40}{\pico\gram} cantilever in an atomic force microscope (AFM). For AFM cantilevers operating at room temperature close to their fundamental resonant frequency in the kilohertz to megahertz range, optical interferometric and beam-deflection detectors of motion are sufficient to resolve the thermal noise force determined by the damping of the cantilever eigenmode in thermal equilibrium with its environment. Operation in high vacuum and at cryogenic temperatures reduces this force noise, improving sensitivity to the point where motion detection becomes the limiting source of noise. In this context the principles of cavity optomechanics may improve the sensitivity of AFM force sensors. Cryogenic AFM further enables the use of superconducting microwave resonators in a cavity optomechanical detection scheme~\cite{Regal2008nanomechanicalbeam, Teufel2009belowsql, Weber2016graphene, Bothner2022kerroptomechanics}. We recently introduced such a sensor based on the electromechanical coupling between surface strain and kinetic inductance of a superconducting nanowire~\cite{Roos2023kimec}. In this paper, we describe in detail the fabrication and characterization methods of these Kinetic Inductance Mechano-Electric Coupling (KIMEC) sensors.

\begin{figure*}
    \centering
    \includegraphics[width=0.9\linewidth]{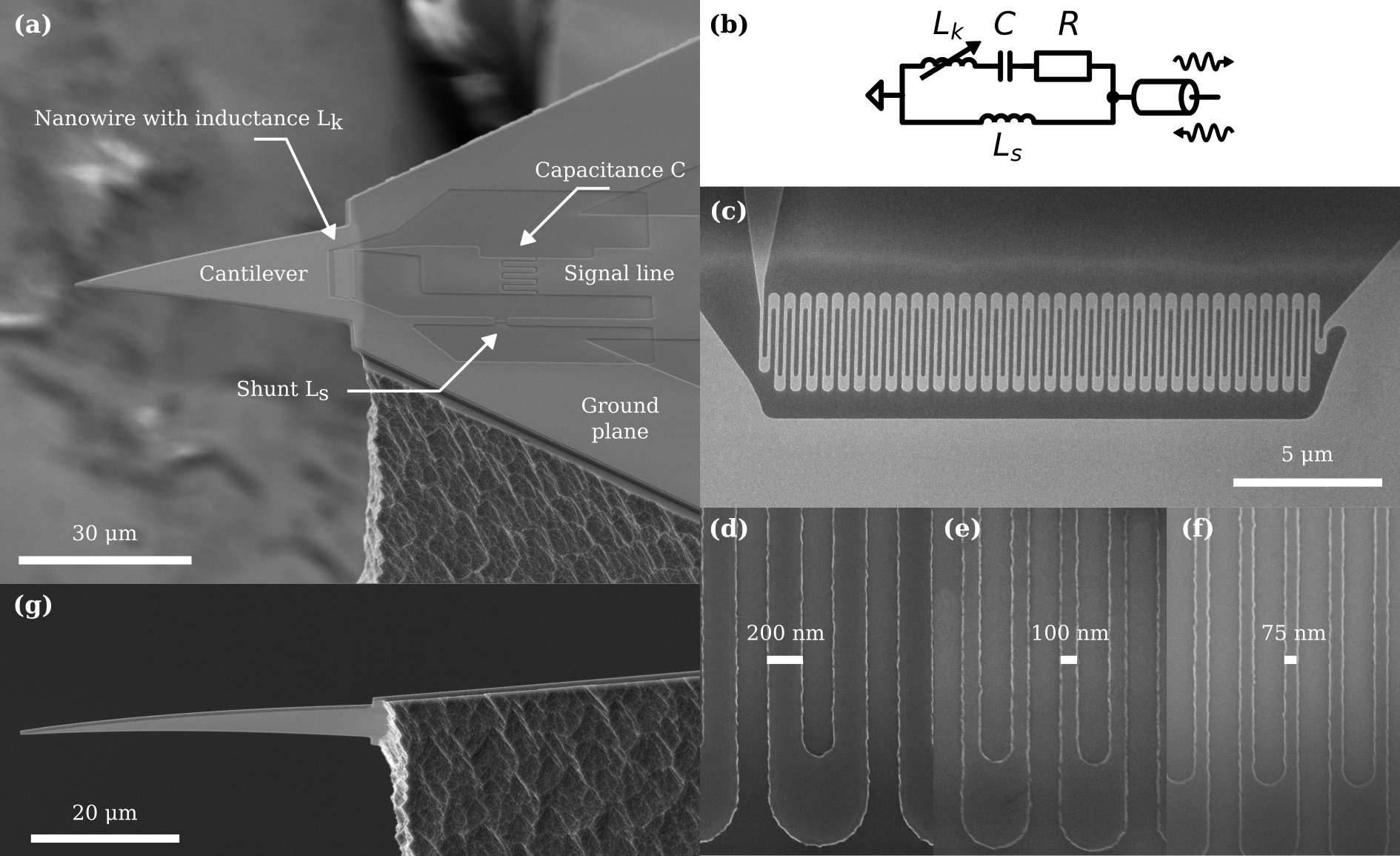}
    \caption{
		(a) A scanning electron microscope (SEM) image of a fabricated sensor seen from an angled topside view. The cantilever is formed from a Si-N plate protruding from and supported by a Si substrate. A thin film of Nb-Ti-N is deposited on top of the Si-N and patterned to form the microwave resonator. A short nanowire in parallel forms a shunt inductance $L_{s}$, affecting the coupling between the resonator and the transmission line. The signal line of a coplanar waveguide is connected to the circuit. 
		(b) Equivalent circuit diagram of the device, where the microwave mode is modeled as a series $R L_{k} C$-circuit in parallel with a shunt inductance $L_{s}$, directly connected to a transmission line and measured in reflection. The resistance $R$ models the internal losses of the microwave resonator. 
		(c) A topside view of the meandering nanowire inductor at the base of the cantilever with kinetic inductance $L_{k}$. The inductor is placed transverse to the base of the released cantilever.
		(d)--(f) SEM images of nanowires from three different devices, showing three different nominal nanowire widths: \SIlist{200;100;75}{\nano\meter}. 
		(g) An SEM image of an underside view of the clamping line of a released cantilever using an isotropic silicon etch. The etch produces an uneven clamping line, affecting the mechanical frequency of the cantilever.
    }
    \label{fig:device_design}
\end{figure*}

A force sensor designed specifically for scanning probe microscopy must have a sharp tip that is readily positioned and scanned over a surface. We operate the sensor near a mechanical resonance with a high-quality factor $Q$ for enhanced responsivity to force. The mechanical resonator is a transducer, converting tip-surface force to changes in its sinusoidal motion as the tip oscillates above the surface with amplitude $A \simeq \SI{1}{\nano\meter}$, the typical range of tip-surface forces. Furthermore, we desire that the stiffness of the mechanical mode $k \simeq \SI{100}{\newton\per\meter}$, the typical change in tip-surface force gradient during oscillation. Although many types of resonators could fulfill these requirements in principle, it is hard to beat the micro-cantilever for ease of fabrication.

To sense force on the tip we need to measure the motion of the mechanical resonator, detecting its deflection from mechanical equilibrium. With AFM, this is typically done by optoelectronic means, using either interferometry~\cite{Rugar1989fiber, Srinivasan2011cantilever, Fogliano2021fiber}, or beam deflection~\cite{Alexander1989optical, Pottier2017thermalnoiseprocessing, Ma2021contactafm}. These optical methods often require delicate in-situ alignment of the detector to the mechanical force transducer. An integrated detector requiring no alignment is highly desirable. Furthermore, we would like the integrated sensor package, i.e. the transducer and detector, to be easily exchangeable as AFM tips are frequently damaged when scanning over unknown surface features.  

Dynamic AFM is typically operated in two alternative modes of scanning feedback: Amplitude Modulation AFM (AM-AFM) and Frequency Modulation AFM (FM-AFM). Both modes, and their many variants, rely on lock-in detection of the motion signal, and in most cases, this signal is at the same frequency as the excitation. Cavity optomechanical detection principles can be used for both AM-AFM and FM-AFM, as well as additional driving and readout schemes. In contrast with optical cavities, superconducting lumped-element microwave resonators easily reach the resolved-sideband regime, where the cavity linewidth is smaller than the mechanical resonance frequency~\cite{Aspelmeyer2014review, Teufel2011sidebandcooling}. Typically, lumped-element microwave resonators are coupled to mechanical motion through a change in capacitance. Here, we detect motion through a change in kinetic inductance.  

Kinetic inductance is an electromechanical phenomenon resulting from Cooper pair mass and the kinetic energy of a supercurrent. It can be orders of magnitude larger than geometric (electromagnetic) inductance in thin films and nanowires made of amorphous superconductors~\cite{Zmuidzinas2012review} and therefore useful in applications that require compact microwave resonators with low loss~\cite{Krantz2019review}, including microwave filters~\cite{Carroll1993} and resonant radiation detectors~\cite{Vissers2015}. Large kinetic inductance also comes with intrinsic nonlinearity, or current dependence of the inductance, which enables low-noise microwave parametric amplification~\cite{Parker2022parametricamplification}. Different materials studied in the literature include niobium nitride (Nb-N)~\cite{Niepce2019kinetic, Anferov2020}, titanium nitride (Ti-N)~\cite{Shearrow2018, Joshi2022}, niobium titanium nitride (Nb-Ti-N)~\cite{Parker2022parametricamplification, Bretz-Sullivan2022kinetic}, or granular aluminum (grAl)~\cite{Maleeva2018, Valenti2019}, wolfram (W)~\cite{Basset2019}, and silicon doped with boron (Si:B)~\cite{Bonnet2022}.

\section{Results and Discussion}

Figure~\ref{fig:device_design} gives an overview of our fabricated sensor, showing the main components. The cantilever is a \SI{600}{\nano\meter} thick Si-N triangular plate released from a much thicker silicon (Si) support, as shown in Figs.~\ref{fig:device_design}(a) and \ref{fig:device_design}(g). Figure~\ref{fig:device_design}(a) shows the microwave resonant circuit with its interdigital capacitor in series with the nanowire inductor. The meandering nanowire is placed at the base of the cantilever as shown in Fig.~\ref{fig:device_design}(c), in the area of largest strain. Details of three different nanowire widths are shown in Figs.~\ref{fig:device_design}(d)-(f). The circuit is measured in reflection, as illustrated in the device schematic in Fig.~\ref{fig:device_design}(b), through a coaxial transmission line that launches to the coplanar waveguide on the sensor chip (not shown). We now consider the sensor design in more detail.

\subsection{Sensor design}
We view the sensors as composed of the cantilever that transduces force to displacement (transducer) and the microwave resonator that detects the displacement (detector). For this reason we separately discuss the mechanical resonator, the microwave resonator and their electromechanical coupling. The coupling of the microwave resonator to a transmission line is also an important design consideration.

\begin{figure}
    \centering
    \includegraphics[width=0.9\linewidth]{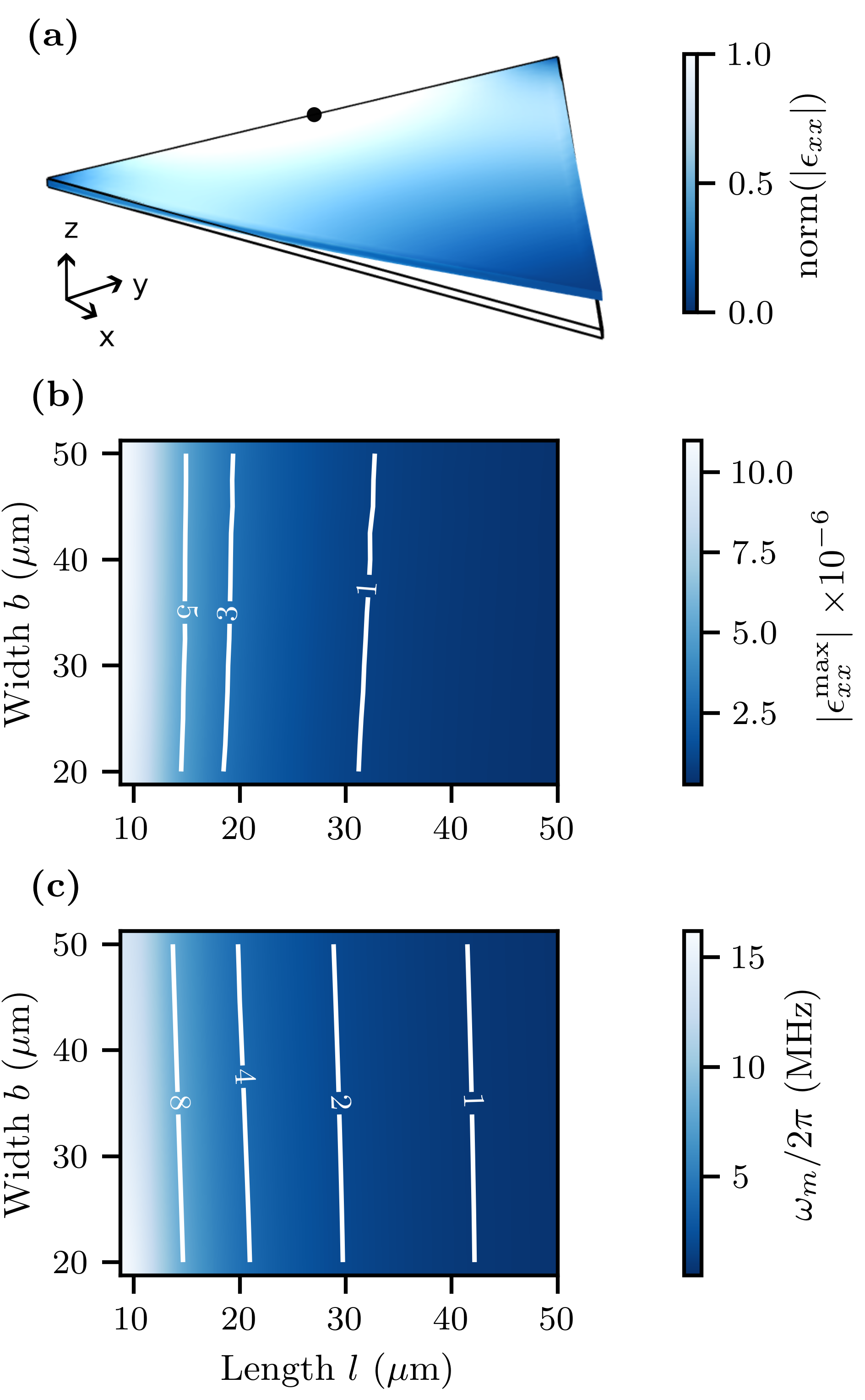}
    \caption{
        (a) The distribution of longitudinal strain at the surface for displacement in the $z$-direction, $\epsilon_{xx}(x,y)$, a dimensionless quantity ($\epsilon = \Delta l / l$). The strain is simulated for a \SI{1}{\nano\meter} displacement at the apex of the triangular Si-N plate in the $z$-direction, assuming a perfect clamp along the base of the plate. A lighter color indicates a larger strain. The displacement of the cantilever is exaggerated for clarity.
        (b) The maximum strain $\vert \epsilon_{xx}^{\mathrm{max}} \vert$ at the point indicated by the dark dot [see panel (a)], as a function of cantilever width $b$ and length $l$.
        (c) The resonant frequency of the cantilever $\omega_{m}$ as a function of its width $b$ and length $l$.
    }
    \label{fig:strain_fem_simulation}
\end{figure}

Our primary goal is high sensitivity to tip-surface forces, given the constraints of the AFM application. If the detector is thermal-noise limited, the sensitivity is given by the power spectral density of fluctuations in force (force noise) $S_{\mathrm{FF}}(\omega)$ which sets a minimum detectable force (signal-to-noise ratio equals one) in a given measurement bandwidth, i.e. signal integration time. The fluctuation-dissipation theorem applied to the harmonic oscillator gives
\begin{equation}
    S_{\mathrm{FF}} = 4 \kB T_{m} \meff \gamma_{m},
    \label{eqn:power_spectral_density_noise_force}
\end{equation}
where $T_{m}$ is the temperature of the mechanical mode and the damping coefficient is expressed as the product of the effective mass $\meff$ and the linewidth $\gamma_{m}$ of the mechanical resonance. Decreasing the temperature and damping improves force sensitivity. It is important to understand the sources of damping for cryogenic AFM, where cantilevers oscillate in vacuum. Force sensitivity will improve with smaller $\meff$ only if mechanisms of damping, such as clamping loss and surface effects, do not increase disproportionately.

The detection scheme imprints the mechanical motion into the field of the microwave resonator, leading to motional sidebands in the measured output microwave field $S_{VV}(\omega)$. The thermal noise force is detected at these sidebands~\cite{Roos2023kimec},
\begin{equation}
    S_{VV}(\omega) = S_{VV}^{\mathrm{NF}} + \alpha n_{c} g_{0}^{2} \frac{\kB T_{m}}{\hbar \omega_{m}} \frac{\gamma_{m}}{(\omega - \omega_{m})^{2} + \gamma_{m}^{2}/4}
    \label{eqn:measured_thermal_mechanical_displacement}
\end{equation}
where $S_{VV}^{\mathrm{NF}}$ is the added noise of the detector, $n_{c}$ is the number of circulating intra-cavity photons in the microwave resonator, $g_{0}$ is the single-photon electromechanical coupling rate, and $\alpha$ is a proportionality factor that depends on excitation power, the resonator parameters and the measurement line. If the motional sideband is larger than the noise floor of the detector, then force sensitivity is set by the properties of the mechanical resonator.

First in the hierarchy of design constraints comes the spring constant $k$, which should match the maximum tip-surface force gradient. For a given $k$ the mechanical resonant frequency is set by $\meff$,
\begin{equation}
    k = \meff \omega_{m}^{2}.
    \label{eqn:spring_constant}
\end{equation}
A larger mechanical resonant frequency gives a larger integration bandwidth for a given mechanical quality factor $Q_{m} = \omega_{m} / \gamma_{m}$. However, increasing $\omega_{m}$ also requires decreasing $\meff$ to maintain $k$. Practically, the limits of $\omega_{m}$ and $\meff$ are set by material choices, fabrication possibilities, and cantilever dimensions. These factors also affect the surface strain at the base of the cantilever where the nanowire is located. 

Second in the hierarchy is the resonant frequency $\omega_{c}$ and linewidth $\kappa$ of the microwave resonator used to detect cantilever motion. A practical consideration is that our multifrequency measurement apparatus works in the frequency band \SIrange{4}{8}{\giga\hertz}. This frequency range constrains the possible values of the circuit's inductance $L$ and capacitance $C$. The relatively small size available from the cantilever dimensions motivates the use of kinetic inductance to achieve a compact lumped-element inductor with negligible stray capacitance. Mattis-Bardeen theory~\cite{Annunziata2010} relates the kinetic inductance of a film with $t$ much less than the London penetration depth, to the normal-state sheet resistance $R_{\square}$ and the superconducting energy gap $\Delta_{0}$. For a conductor of length $\ell$ and width $w$, the kinetic inductance is
\begin{equation}
    L_{k} = \left( \frac{\ell}{w} \right) \frac{\hbar R_{\square} }{\pi \Delta_{0}} = \left( \frac{\ell}{w} \right) \Lks,
    \label{eqn:kinetic_per_square}
\end{equation}
where $\Lks$ is the kinetic inductance per square. From Eqn.~\eqref{eqn:kinetic_per_square}, we require a superconducting film with large normal state resistance in geometry with many squares, i.e. a long and narrow strip. For a given total inductance $L$, the meandering structure allows for a physically compact shape. A larger $L$ also leads to a larger maximum number of intra-cavity photons $n_{c}$ for a given current $I$, 
\begin{equation}
    n_{c} = \frac{\Lks}{2 \hbar \omega_{c}} \left(\frac{\ell}{w} \right) I^{2},
    \label{eqn:intracavity_photons_for_inductance}
\end{equation}
where the total current through the inductor is limited by the critical current $I_{c}$, also set by material choice and physical dimensions. From Eqn.~\eqref{eqn:intracavity_photons_for_inductance}, we can increase $n_{c}$ for a given material by, again, making the nanowire longer or narrower, or by increasing $\Lks$ with a thinner film. However, longer and narrower nanowires decrease the participation ratio, i.e. the fraction of the nanowire that experiences strain, and therefore $g_{0}$. The nanowire can instead be made more compact by decreasing the width and adjusting the length so that the same number of squares experience strain. Conversely, reducing the cross-sectional area of the nanowire reduces $I_{c}$ and decreases the maximum $n_{c}$ possible before undesirable nonlinear effects become significant. A nonlinear microwave mode is not part of the standard electromechanical formulation, complicating the analysis for, e.g., the phase-sensitive detection scheme.

\begin{figure*}
    \centering
    \includegraphics[width=0.9\linewidth]{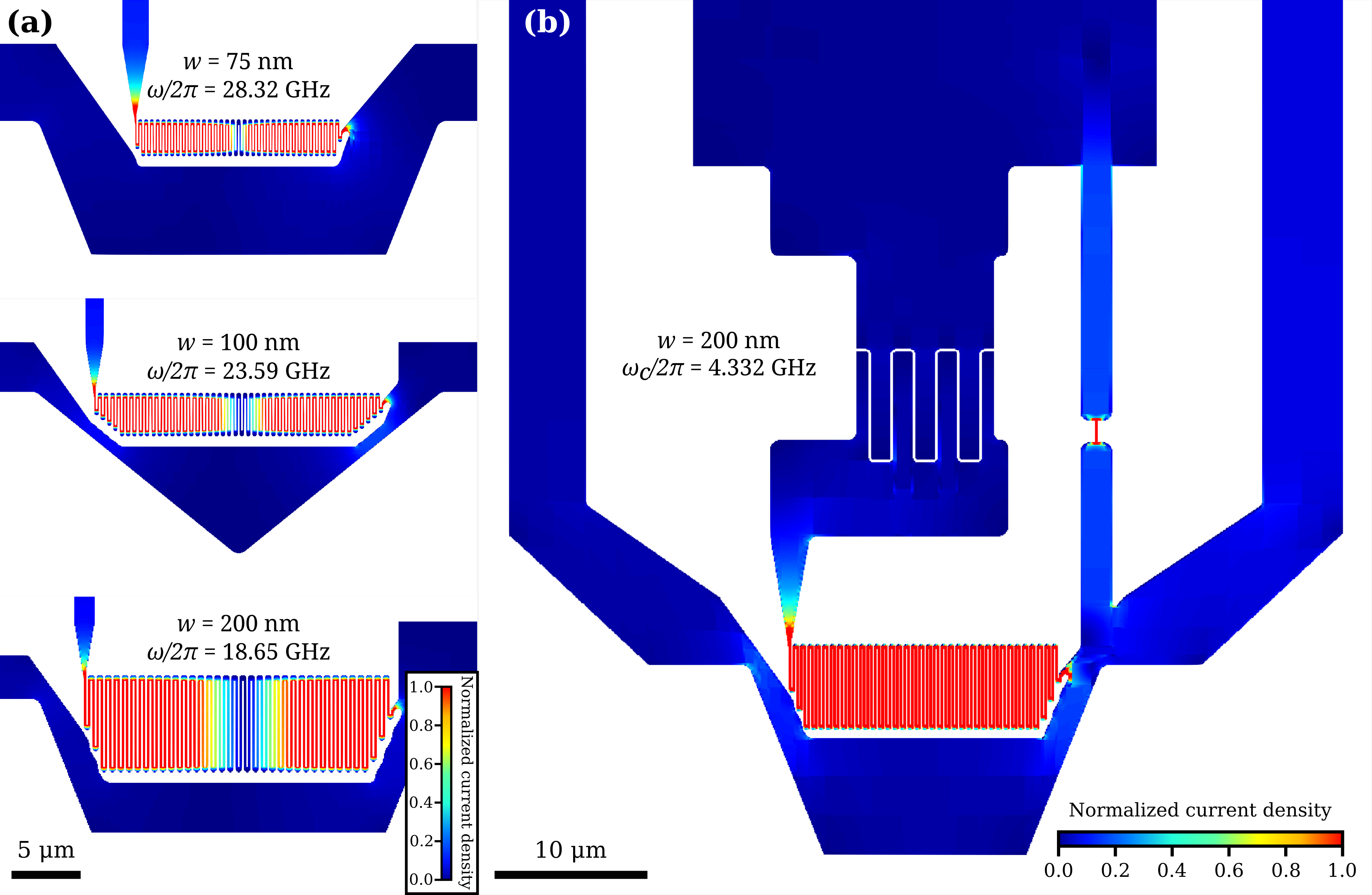}
    \caption{
        (a) Microwave simulations of the normalized current density of the first electromagnetic eigenfrequency $\omega$ of the meandering nanowire, for widths $w$ = \SIlist{75;100;200}{\nano\meter} and using $L_{k, \square}$ = \SI{36}{\pico\henry\per\sq}, with the current node at the center of the nanowire. The frequencies land in the range \SIrange{18}{28}{\giga\hertz}.
        (b) Simulation of the first resonant mode of the entire structure, using nanowire width $w$ = \SI{200}{\nano\meter}. On resonance, the current density is uniformly distributed in the meandering nanowire, behaving as a lumped-element inductor.
    }
    \label{fig:em_simulation}
\end{figure*}

Third in the hierarchy is the value of the single-photon electromechanical coupling strength,
\begin{equation}
    g_{0} = \frac{\partial \omega_{c}}{\partial z} \sqrt{\frac{\hbar}{2 \meff \omega_{m}}} \equiv G \zzpf,
    \label{eqn:vacuum_optomechanical_coupling_rate}
\end{equation}
characterizing the microwave frequency shift $G$ per zero-point motion (in the $z$-direction) of the mechanical mode $\zzpf$. The shift will depend as
\begin{equation}
    G = \frac{\partial \omega_{c}}{\partial L} \frac{\partial L}{\partial z} = -\frac{1}{2} \frac{\omega_{c}}{L} \frac{\partial L}{\partial z},
    \label{eqn:frequency_shift}
\end{equation}
where the last term requires a microscopic theory of the effect of strain on the kinetic inductance of the nanowire. In our design, a sufficient requirement on $g_{0}$ is that the motional sidebands due to the thermal noise force can be resolved above the noise floor of the detector. We prioritize a larger critical current of the nanowire to compensate a potentially smaller $g_{0}$.

\begin{figure*}
    \centering
    \includegraphics[width=0.9\linewidth]{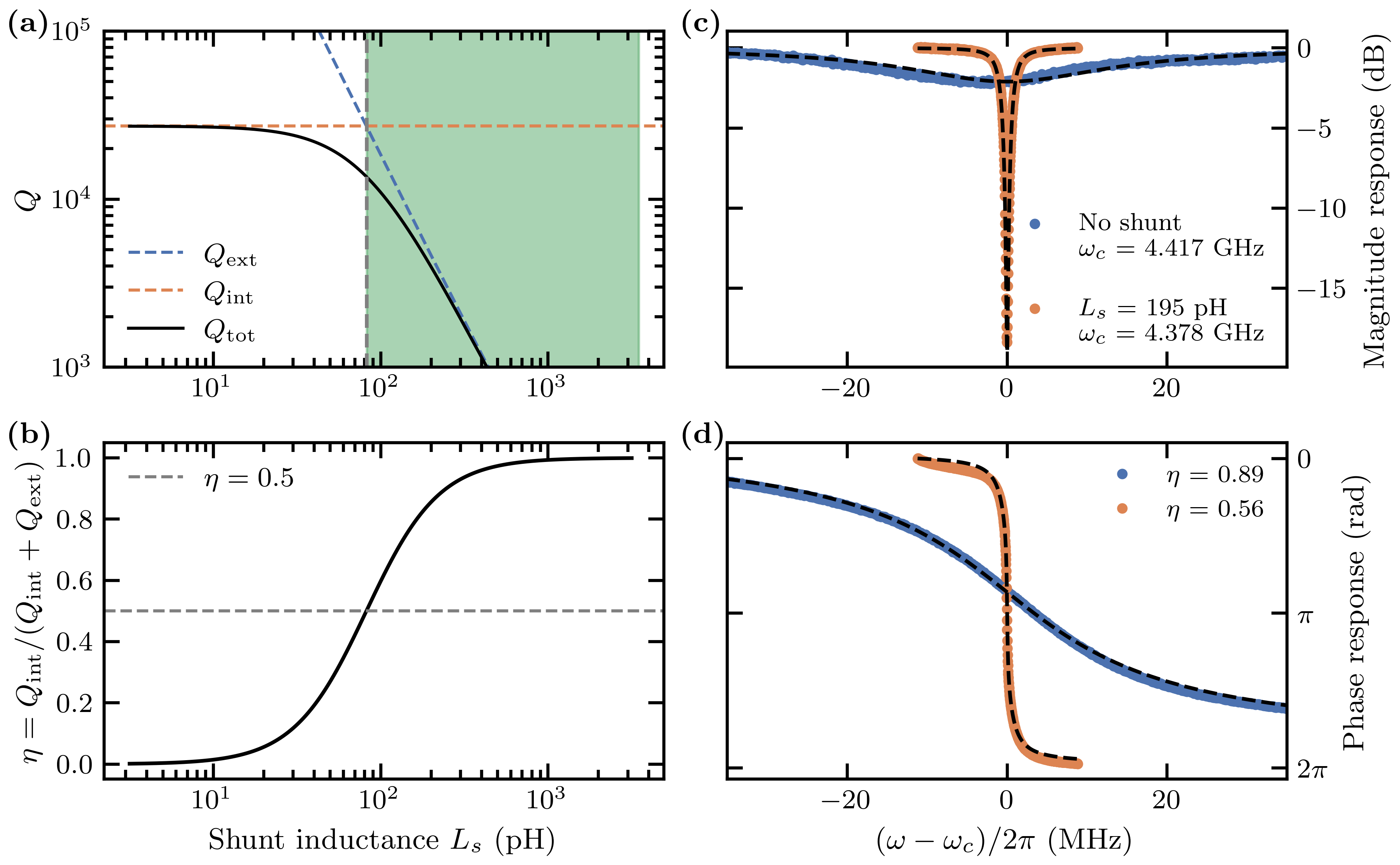}
    \caption{
        (a) Simulated quality factors $\Qtot$, $\Qext$ and $\Qint$, and (b) coupling parameter $\eta = \Qint/(\Qint + \Qext)$ of the circuit in Fig.~\ref{fig:device_design}(b) as a function of the shunt inductance $L_{s}$ for values $R$ = \SI{0.1}{\ohm}, $C$ = \SI{13.5}{\femto\farad} and $L_{k}$ = \SI{100}{\nano\henry}. The circuit is under-coupled to the transmission line with impedance $Z_{0}$ = \SI{50}{\ohm} for low values of $L_{s}$. The external losses increase as the inductance of the shunt $L_{s}$ increases. The circuit is first critically coupled for $L_{s} = \sqrt{Z_{0} R / \omega_{c}^{2}}$ and then over-coupled for $L_{s} > \sqrt{Z_{0} R / \omega_{c}^{2}}$, indicated by the shaded area. The phase response is sharper when the circuit is slightly over-coupled. The line in panel (b) is given by Eqn.~\eqref{eqn:eta_expression}. 
        (c) Measured magnitude and (d) phase response of two samples, with and without a shunt inductance (dots). The dotted line is the fit of the model to the data of an ideal $RLC$-circuit for the data without a shunt and a $RLC$-circuit in parallel with a shunt $L_{s}$ for the shunted sample. The lack of a shunt is equivalent to $L_{s} \rightarrow \infty$. Both samples are designed to be largely over-coupled to the transmission line. The sample with the shunt displays a sharper phase response and a larger dip in the magnitude response, showing that it is closer to the critical coupling than the sample without a shunt. The resonant frequencies for the samples are $\omega_{c} / 2 \pi$ = \SI{4.482}{\giga\hertz} (no shunt) and \SI{4.378}{\giga\hertz} (shunt).
    }
    \label{fig:shunt_inductance}
\end{figure*}

For excitation and read-out, the microwave resonator is coupled to the transmission line with impedance $Z_{0}$ = \SI{50}{\ohm}, giving a measured (total) quality factor $\Qtot$,
\begin{equation}
    \frac{1}{\Qtot} = \frac{1}{\Qint} + \frac{1}{\Qext},
    \label{eqn:total_quality_factor}
\end{equation}
where $\Qint$ and $\Qext$ are the internal and external quality factors of the microwave resonator. We introduce a shunt inductor with inductance $L_{s}$ in order to tune $\Qext$ and, therefore, the coupling parameter $\eta$ of the circuit to the transmission line. The coupling parameter is given by the ratio of external losses, i.e. useful signal, to total cavity losses, or equivalently,
\begin{equation}
    \eta(L_{s}) = \frac{\Qint}{\Qint + \Qext(L_{s})} = \left( 1 + \frac{Z_{0} R}{\omega^2 L_{s}^2} \right)^{-1},
    \label{eqn:eta_expression}
\end{equation}
where $R$ models the losses of the microwave resonator. In the under-coupled case $\eta < 0.5$, the internal losses of the resonator dominate, while in the over-coupled case $\eta > 0.5$, the losses in the transmission line dominate. A large $\Qtot$ and $\eta$ are desirable to maximize the output signal. A smaller shunt inductance increases $\Qext$. However, the total quality factor is bounded by the internal losses. To decrease $\eta$ while remaining overcoupled, the shunt inductor $L_{s}$ should satisfy $\SI{50}{\ohm} > i \omega_{c} L_{s} > R$.

As a final consideration, the temperature range when operating low-temperature AFMs is typically around \SI{1}{\kelvin}. For temperatures closer to $T_{c}$, the internal losses of the microwave circuit increases due to increasing quasiparticle population, leading us to select Nb-Ti-N with its high (bulk) critical temperature $T_{c} \approx \SI{14}{\kelvin}$.

\subsection{Mechanical simulation}

Several considerations determine the design of the Si-N cantilever in the scanning force sensor. The Si-N plate thickness is fixed when fabricating a wafer of sensor chips. We design the cantilever's planeview dimensions to achieve $\omega_{m} /2 \pi$ in the range \SIrange{0.5}{10}{\mega\hertz}, corresponding to mechanical spring constant $k$ in the range \SIrange{2}{160}{\newton\per\metre} for typical device parameters. The wide frequency range allows us to fabricate devices working in either the sideband-resolved or sideband-unresolved regime. For the given thickness, we simulate the eigenfrequencies of the cantilever using the finite-element method (FEM) implemented in COMSOL Multiphysics~\cite{comsol}, with the boundary condition of a perfectly rigid clamp along the line where the plate meets the Si substrate.

The FEM model gives the distribution of strain at the surface. Figure~\ref{fig:strain_fem_simulation}(a) shows the distribution of longitudinal strain $\epsilon_{xx} (x,y)$ for the fundamental bending mode of interest. The strain is normalized to its maximum value at the center of the clamping line. Figure~\ref{fig:strain_fem_simulation}(b) displays this maximum value of the surface strain as a function of the length $l$ and width $b$ of the triangular plate calculated for a \SI{1}{\nano\meter} $z$-displacement at the apex of the triangle, a typical tip displacement for measuring surface forces in AFM.

We place the nanowire inductor in this region of maximum longitudinal surface strain, with the long segments of the nanowire oriented parallel to the $x$-axis. The nanowire will thus experience compression or tension as the tip deflects in positive or negative $z$-direction, respectively. We can then vary $b$ and $l$ of the triangular cantilever to increase the strain for a given deflection and to achieve the desired mechanical resonant frequency $\omega_{m}$, as shown in Fig.~\ref{fig:strain_fem_simulation}(c). We see that the length of the cantilever is the main factor affecting the strain and the resonant frequency. The width $b$ must also be adjusted to accommodate a meandering nanowire with a total length $\ell$ large enough to realize the desired kinetic inductance. To understand this constraint, we turn to electromagnetic simulations.

\subsection{Electromagnetic simulation}\label{sec:em_design}

In this work, we explore thin-film nanowires of width $w$ = \SIlist{75;100;200}{\nano\meter} that we can make with a high degree of uniformity using electron-beam lithography and reactive-ion etching. We simulate the electromagnetic response of the meandering nanowire inductors using Sonnet, a quasi-3D electromagnetic simulator~\cite{sonnet} that has the feature of including sheet kinetic inductance $L_{k, \square}$. We begin by simulating the meandering inductor itself to find the lowest-frequency eigenmode. We desire that this eigenfrequency falls well above the target frequency of our resonator $\sim$~\SI{5}{\giga\hertz} so that we may, to a good approximation, treat the meandering nanowire as a lumped-element inductor. Figure~\ref{fig:em_simulation}(a) shows simulations of the current distribution of a typical inductor for all three nanowire widths at their lowest eigenfrequency in the range \SIrange{18}{28}{\giga\hertz}, where we see the current node located in the center of the meander. Figure~\ref{fig:em_simulation}(b) shows the microwave simulation of the entire circuit, including the shunt inductance formed from a short nanowire. At the lower resonant frequency of the inductor and series capacitor, the current is uniformly distributed inside the meandering nanowire, confirming that it behaves as a lumped-element inductor. We also see that on resonance, the current in the impedance-transforming shunt inductor reaches a similar magnitude to that in the meandering inductor.

As discussed previously, we can further increase the total quality factor $\Qtot \propto \sqrt{L/C}$ by increasing the total inductance $L$, either geometrically, i.e. making the nanowire longer or by increasing the kinetic inductance per square $\Lks$, i.e. making the film thinner or the nanowire narrower. If we arbitrarily increase the total length $\ell$ of the nanowire, the parasitic capacitance of the meandering structure will eventually become significant enough to decrease the first eigenmode frequency to the point that it can no longer be treated as a lumped-element inductor. Additionally, to maintain the resonant frequency in the band \SIrange{4}{8}{\giga\hertz}, an increasing $L$ must be matched by a decreasing capacitance $C$, which cannot be made too small in relation to the parasitic capacitance. The measured parameters of our devices given below represent an adequate trade-off between these design considerations.

The shunting inductance forms a part of the circuit design. Figure~\ref{fig:shunt_inductance}(a) shows the simulated total quality factor $\Qtot$ and Fig.~\ref{fig:shunt_inductance}(b) shows the coupling parameters as a function of the shunt inductance, using typical circuit parameters for our devices. Figures~\ref{fig:shunt_inductance}(c) and \ref{fig:shunt_inductance}(d) display the measured magnitude and phase response of two nominally identical devices, both with nanowire width $w$ = \SI{200}{\nano\meter}, where one device has the shunt inductance and the other does not. For a shunt with inductance $L_{s}$ = \SI{195}{\pH}, we increase $\Qext$ by a factor of roughly twenty at the cost of a slight reduction in $\eta$ while remaining over-coupled.

\begin{figure*}
    \centering
    \includegraphics[width=0.9\linewidth]{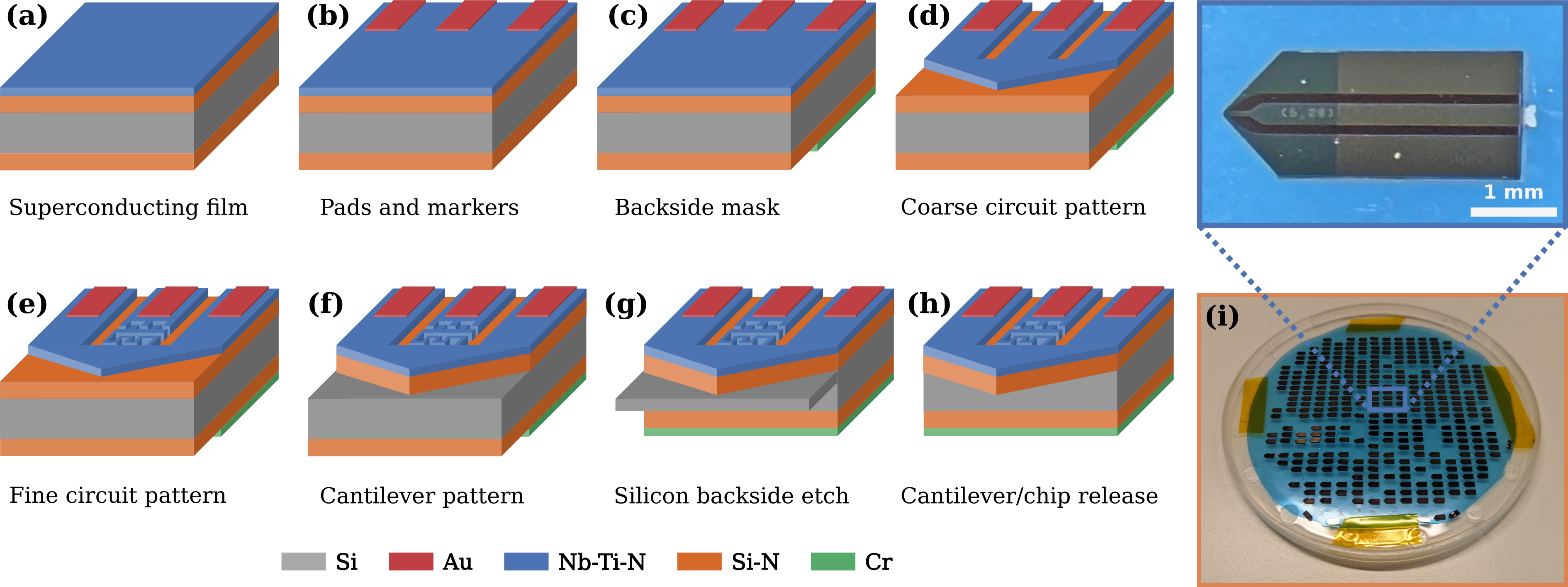}
    \caption{
        (a)-(h) Illustration of the main fabrication steps for one device (not to scale). Details of each step are given in the main text. (i) Photograph of a 100-mm wafer after sensor release, frame removal, and removal of broken chips, with remaining chips attached to the wafer tape. The inset is an optical image of one chip. The cavity and cantilever, too small to be seen in this image, are located at the left-pointed end of the chip 
    }
    \label{fig:fabrication_steps}
\end{figure*}

\subsection{Design summary}\label{sec:design_summary}

The narrative above attempts to convey the complex interplay between the mechanical and electrical considerations when designing the force sensor. Some important constraints are the cavity resonant frequency $\omega_{c} / 2 \pi \sim$~\SI{5}{\giga\hertz}, which must be inside the band \SIrange{4}{8}{\giga\hertz} of our digital multifrequency lock-in measurement system. Higher frequencies are possible, but the cost of such equipment increases steeply with frequency. Our designs also started with a single thickness of the Si-N plate that forms the cantilever, which is natural as all chips must be fabricated on the same Si-N layer. By increasing the thickness of the Si-N plate, we increase the surface strain for the same curvature of the cantilever, giving larger $G = \frac{\partial \omega_{c}}{\partial L} \frac{\partial L}{\partial z}$, which also increases the stiffness of the bending mode. We can compensate for the latter by increasing the length and reducing the width of the triangular plate. Eventually, we run out of space to accommodate the kinetic inductor. However, the inductor's footprint can be reduced by increasing its sheet kinetic inductance using a thinner superconducting film. Another option is to work with smaller inductance, which would require increasing the capacitance of the resonator to keep the resonant frequency constant. However, as previously mentioned, $L$ or $C$ cannot be arbitrarily large or small without considering parasitic inductance or capacitance.

We do not claim to have achieved an optimal design and clearly there is plenty of room to explore variations. We have discussed the importance of $G$ as a figure-of-merit. This parameter provides the calibration of cantilever tip motion, which is an important part in design optimization. A good design must also be fabricated at the wafer scale with a reasonable number of process steps and at a reasonable cost and uniformity.

\subsection{Fabrication}\label{sec:fabrication}

The main fabrication steps are illustrated in Fig.~\ref{fig:fabrication_steps}. The fabrication produces around 400 chips per wafer with a yield of about \SI{80}{\percent}. We start with double-side polished Si wafer, \SI{100}{\milli\meter} diameter and \SI{525}{\micro\meter} thick, coated on both sides with \SI{600}{\nano\meter} low-stress ($<$ \SI{100}{\mega\pascal}) Si-N films. Sensor chips are \SI{1.6}{\milli\meter} by \SI{3.4}{\milli\meter}, about the size of a standard AFM cantilever chip. The steps are as follows:

\begin{figure*}
    \centering
    \includegraphics[width=0.9\linewidth]{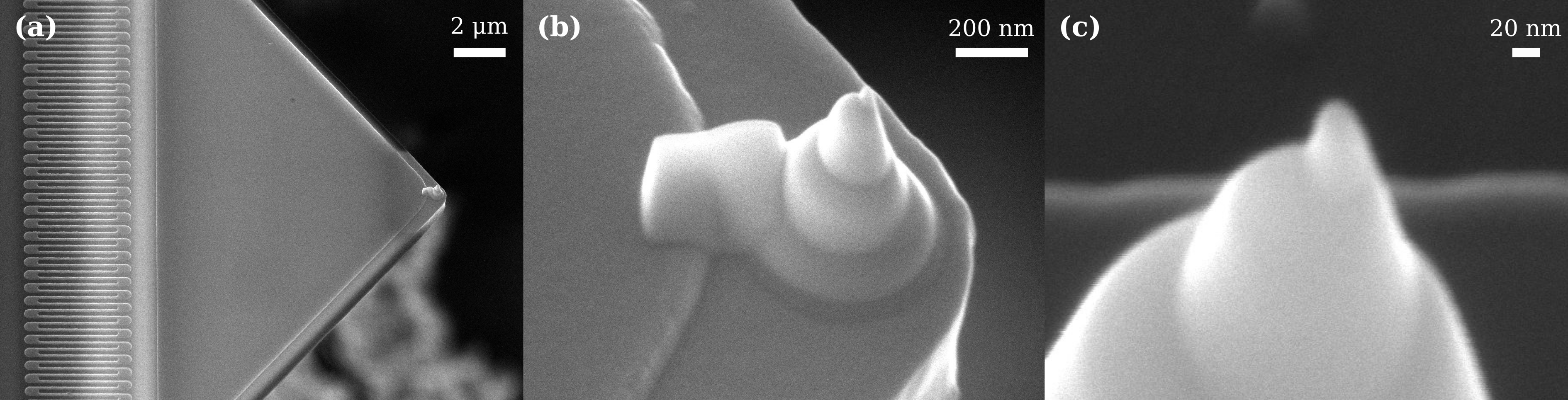}
    \caption{
        (a) An SEM image of a top-tilted view of an electron-beam deposited platinum tip on the released Si-N cantilever of one chip. The tip has total height in the range \SIrange{1}{2}{\micro\meter}.
        (b) An SEM image of the reverse cone structure, deposited using multiple layers of platinum. Note the larger connection to the Nb-Ti-N thin film to the left of the tip, ensuring an ohmic contact with the ground plane of the chip.
        (c) An SEM-image of a deposited tip, showing a radius of curvature smaller than \SI{10}{\nano\meter}.
    }
    \label{fig:tip_deposition}
\end{figure*}

\begin{itemize}
    \item[(a)] \textbf{Superconducting film.} We first deposit a \SI{15}{\nano\meter} thick thin film of superconducting \ce{Nb_{60}Ti_{40}N} by reactive co-sputtering from separate niobium and titanium targets~\cite{zichi2019nbitin} in an ATC2200 from AJA International Inc., with a deposition rate of roughly \SI{3}{\nano\meter\per\minute}.
    
    \item[(b)] \textbf{Pads and markers.} A lift-off process defines the gold contact pads and alignment marks. We spin a \SI{400}{\nano\meter} thick photoresist (maN1407), bake on a hotplate at \SI{100}{\celsius} for \SI{60}{\second} and expose with a dose of \SI{450}{\milli\joule\per\square\centi\meter} using an MLA150 from Heidelberg Instruments. We develop the pattern in maD533s for roughly \SI{45}{\second} and then deposit \SI{10}{\nano\meter} chromium (Cr) and \SI{40}{\nano\meter} of gold by electron-beam evaporation in an Auto306 from Edwards Vacuum. Lift-off in mrREM700 removes the resist mask and the patterned wafer is ultrasonically cleaned and rinsed with IPA.
    
    \item[(c)] \textbf{Backside mask.} Before fabricating the chromium etch mask on the backside, we first protect the wafer's front side with a thin PMMA layer. We then define the lift-off mask on the wafer back side by spinning a \SI{400}{\nano\meter} thick photoresist (maN1407), baking at \SI{100}{\celsius} on a hotplate for \SI{60}{\second}. We expose the pattern with a dose of \SI{450}{\milli\joule\per\square\centi\meter}, aligning to the markers on the front, and we develop in maD533s for roughly \SI{45}{\second}. A subsequent short soft-ashing step in a Plasmalab 80 ICP65 from Oxford Instruments removes residual resist and improves the adhesion of the following \SI{150}{\nano\meter} deposition of chromium with electron-beam evaporation in the Auto306 from Edwards. After the lift-off in mrREM700, we also strip the protective PMMA layer on the front side with AR600--71, and we clean the wafer in isopropanol (IPA).
    
    \item[(d)] \textbf{Coarse circuit pattern.} A layer of photolithography defines the coarse circuit features in the superconducting film, such as the coplanar waveguide, ground planes, and signal line. We use the same recipe as in steps (b) and (c) to define a resist etch mask and after development, we transfer the pattern into the superconducting film with a \ce{CF_{4}}/\ce{O_{2}} reactive-ion etch (RIE) process in a Plasmapro 100 ICP300 from Oxford Instruments, with an etch rate of roughly \SI{8}{\nano\meter\per\minute}.
    
    \item[(e)] \textbf{Fine circuit pattern.} Electron-beam lithography defines the finer structures, such as the meandering nanowire inductor, the shunt inductor, and the interdigital gap of the capacitor. We first spin a thin layer of an adhesion promoter (AR 300--80), before spinning a roughly \SI{170}{\nano\meter} thick layer of the electron-beam resist ARP--6200--09 (CSAR 09), baking at \SI{150}{\celsius} for \SI{1}{\minute}. We expose with a dose of \SI{110}{\micro\coulomb\per\square\centi\meter} in a Voyager EBL system from Raith Nanofabrication, and etch the Nb-Ti-N film using the same \ce{CF_{4}}/\ce{O_{2}} RIE-process as in step (d). In our design, we vary the widths ($w$ = \SIlist{75;100;200}{\nano\meter}) of the nanowires across the wafer, adjusting the total number of squares (total inductance) and the capacitor to obtain a resonant frequency $\omega_{c} / 2 \pi \sim$~\SI{4.5}{\giga\hertz}, see Fig.~\ref{fig:device_design}(d)--(f).

    \item[(f)] \textbf{Cantilever pattern.} Photolithography defines the chip and cantilever. We spin a \SI{1.7}{\micro\meter} thick photoresist maP1225, bake at \SI{105}{\celsius} for \SI{2}{\minute}. We then expose with dose \SI{300}{\milli\joule\per\square\centi\meter} in the MLA150, and develop in maD331 for \SI{45}{\second}. We etch through the Si-N layer using a \ce{CHF_{3}}/\ce{SF_{6}} process with an etch rate of roughly \SI{100}{\nano\meter\per\minute} in the Plasmapro 100 ICP300.

    \item[(g)] \textbf{Backside through-etch.} Before etching through the back side of the wafer, we first spin a protective positive resist on the front side and pattern an opening, or a ``trench'', around the chip that we will use to complete the etch once a larger portion of the wafer has been etched through from the back. We design the trench so that all cantilevers on the wafer are released at the same time, irrespective of their length. To this end we spin a roughly \SI{2.2}{\micro\meter} thick layer of photoresist (maP1225), bake it at \SI{105}{\celsius} for \SI{3}{\minute}, expose with dose \SI{550}{\micro\coulomb\per\square\centi\meter} and develop it in maD331 for \SI{60}{\second}. With the front side of the wafer protected, we flip over the wafer and etch through the Si-N using the same \ce{CHF_{3}}/\ce{SF_{6}} process as in step (f) and the etch mask defined in step (c). We then use a Bosch process to etch through most of the Si substrate (approximately \SI{450}{\micro\meter} deep) with an etch rate of roughly \SI{6}{\micro\meter\per\minute}. This results in all the samples on the wafer being supported by a thin layer of silicon close to the top side.
    
    \item[(h)] \textbf{Release and cantilever under-etch.}  A simple and fast method of release uses an isotropic dry-etch that completes the chip's release from the wafer and removes the unwanted silicon support underneath the silicon nitride cantilever. We etch the silicon through the ``trench'' defined in step (g), from the topside with a short Bosch etch, followed by an isotropic etch that undercuts the cantilever. We use an \ce{SF_{6}}/\ce{O_{2}} RIE-process in the Plasmapro 100 ICP300 with lateral etch rate of \SI{10}{\micro\meter\per\minute}. The isotropic etch results in an uneven clamping line, as shown in Fig.~\ref{fig:device_design}(g), leading to variations in the mechanical resonant frequency
\end{itemize}
Before the final etch and release step, we apply a Nitto Semiconductor Wafer Tape to the backside of the wafer, holding it together during release. After release, the wafer is cleaned in mrREM7000 and IPA, and the individual sensor chips are separated from the wafer in a single step by lifting away the outer frame, with the chips remaining on the wafer tape, as shown in Fig.~\ref{fig:fabrication_steps}(i).

We tested an alternative method to release the cantilever using a wet-etch in potassium hydroxide (KOH). The wet-etch has a high selectivity between silicon and silicon nitride, and KOH etches silicon with different rates in the \hkl<100> and \hkl<111> crystalline directions. With proper orientation of the cantilever mask to the crystalline axes of the wafer one can etch under the triangular silicon nitride plate and form a very straight clamping line to the Si substrate. However, the KOH etch is slow compared to the isotropic RIE process and attacks the Nb-Ti-N superconducting film. An additional lithography step was needed to protect the superconducting circuit with a mask consisting of a \SI{190}{\nano\meter} thick layer of Cr and PMMA. After the release, we strip the PMMA and Cr layer, while taking care to not break the cantilevers. The difficulties associated with using KOH lead us to prefer the dry-etch described in step (h).

\subsection{Tip deposition}

In scanning probe microscopy (SPM) the tip plays a fundamental role in the achievable lateral resolution of the image. The focused electron-beam induced deposition (FEBID)~\cite{utke2012nanofabrication} technique has been adapted to fabricate tips for SPM, for example to enhance commercial platinum-iridium alloy (Pt:Ir) coated conductive tip~\cite{brown2013electrically}, or to realize laterally grown high-aspect ratio nanopillars~\cite{beard2011fabrication}. We realize sharp, vertically grown conductive tips at the apex of the Si-N cantilever using FEBID with a Pt precursor gas. Figure~\ref{fig:tip_deposition} shows the resulting structure. We obtain the conical shape by stacking multiple depositions with different radii to achieve a total tip height in the range \SIrange{1}{2}{\micro\meter}. This conical structure gives added rigidity to lateral forces while scanning. We form a sharp tip at the apex of the cone by exposing a circular area with a diameter of \SI{10}{\nano\meter}, which is smaller than the nominal electron-beam spot size, and by setting the deposition height to \SI{10}{\micro\meter}. Defocusing of the electron spot during vertical growth naturally forms a narrowing conical structure. At the apex of this cone, we routinely achieve a curvature radius of less than \SI{10}{\nano\meter}, as verified by the SEM image in Fig.~\ref{fig:tip_deposition}(c). Finally, we deposit a thin strip to connect the base of the cone to the Nb-Ti-N film which is the ground plane of the microwave circuit. This feature enables the measurement of the tunneling current between the tip (grounded) when a DC bias is applied to a conductive sample surface. Scanning tunneling microscopy (STM) operation was verified both at room temperature and in a cryogenic environment. Thus, the deposited material is suitably conductive for STM and various electrostatic AFM techniques that require applying a low-frequency voltage to the tip.

\begin{figure}
    \centering
    \includegraphics[width=0.9\linewidth]{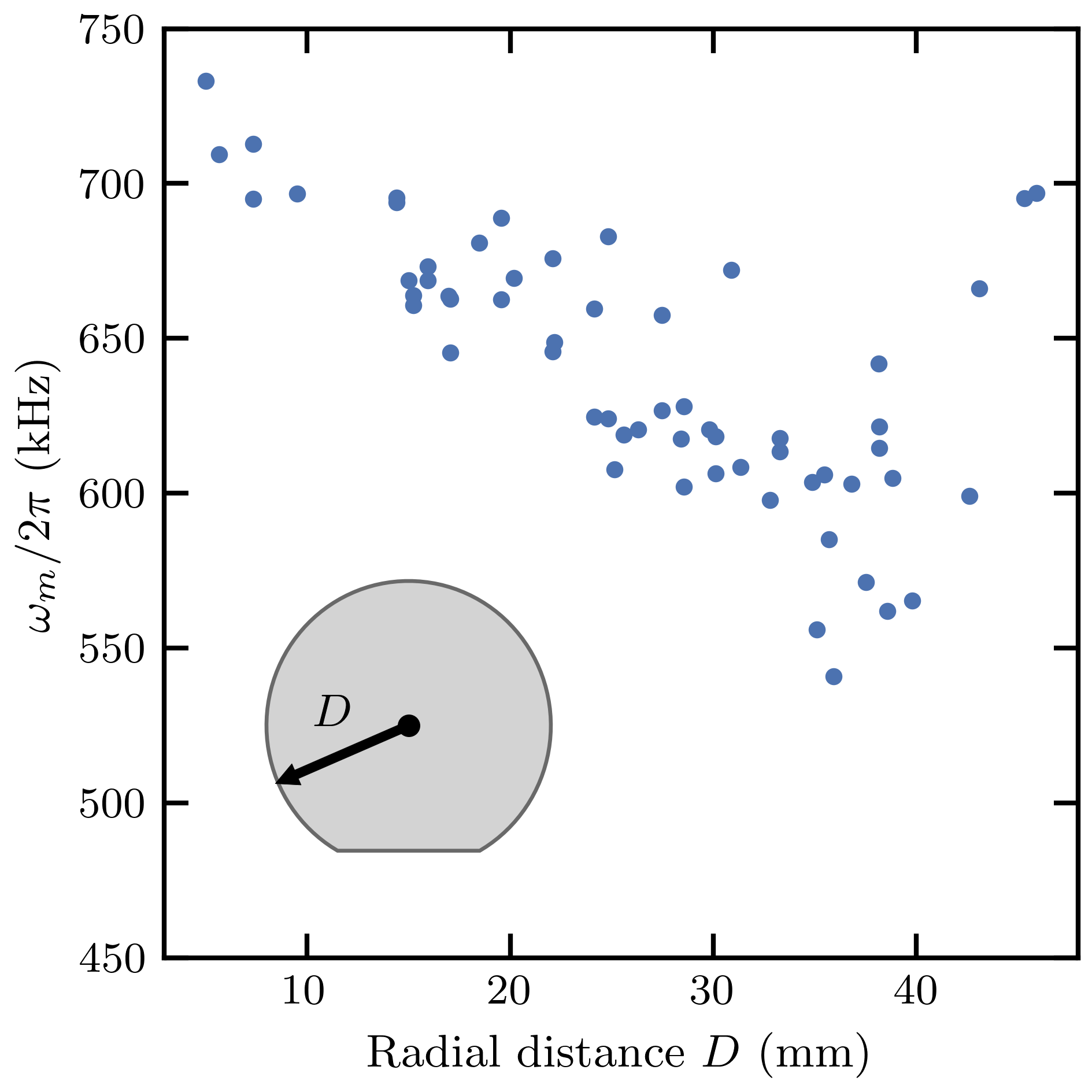}
    \caption{
        Mechanical resonant frequency $\omega_{m}$ of a long cantilever as a function of radial distance $D$ from the center of the wafer (illustrated as an inset). The resonant frequency decreases with the radial distance and with increased spread. Finite-element method simulations give an expected resonant frequency of \SI{700}{\kilo\hertz}.
    }
    \label{fig:mechanical_resonances}
\end{figure}

\subsection{Mechanical mode}\label{sec:mechanical_characterization}

Our chips were the same size as an AFM cantilever chip, making it easy to load them into commercial AFM for mechanical characterization of the cantilever. For the design with nominal cantilever width \SI{40}{\micro\meter}, length \SI{50}{\micro\meter} and thickness \SI{600}{\nano\meter}, the optical lever detector in our AFM had sufficient bandwidth to detect the fundamental bending mode. We measured 60 chips from one wafer, detecting the thermal fluctuations at room temperature in ambient conditions and fitting them to a Lorentzian lineshape. We found that $\omega_{m}$ decreases with the increasing radial distance $D$ from the center of the wafer, as shown in Fig.~\ref{fig:mechanical_resonances}, probably due to non-uniform etching conditions across the wafer. To some extent, one could change the mask design and adjust the dimensions of the cantilever to compensate for this effect. Using the mean value of \SI{641 \pm 42}{\kilo\hertz} and adjusting the Young's modulus of our Si-N plate to \SI{208}{\giga\pascal}, we find good agreement between the mechanical simulation and experiment.

\begin{figure*}
    \centering
    \includegraphics[width=0.9\linewidth]{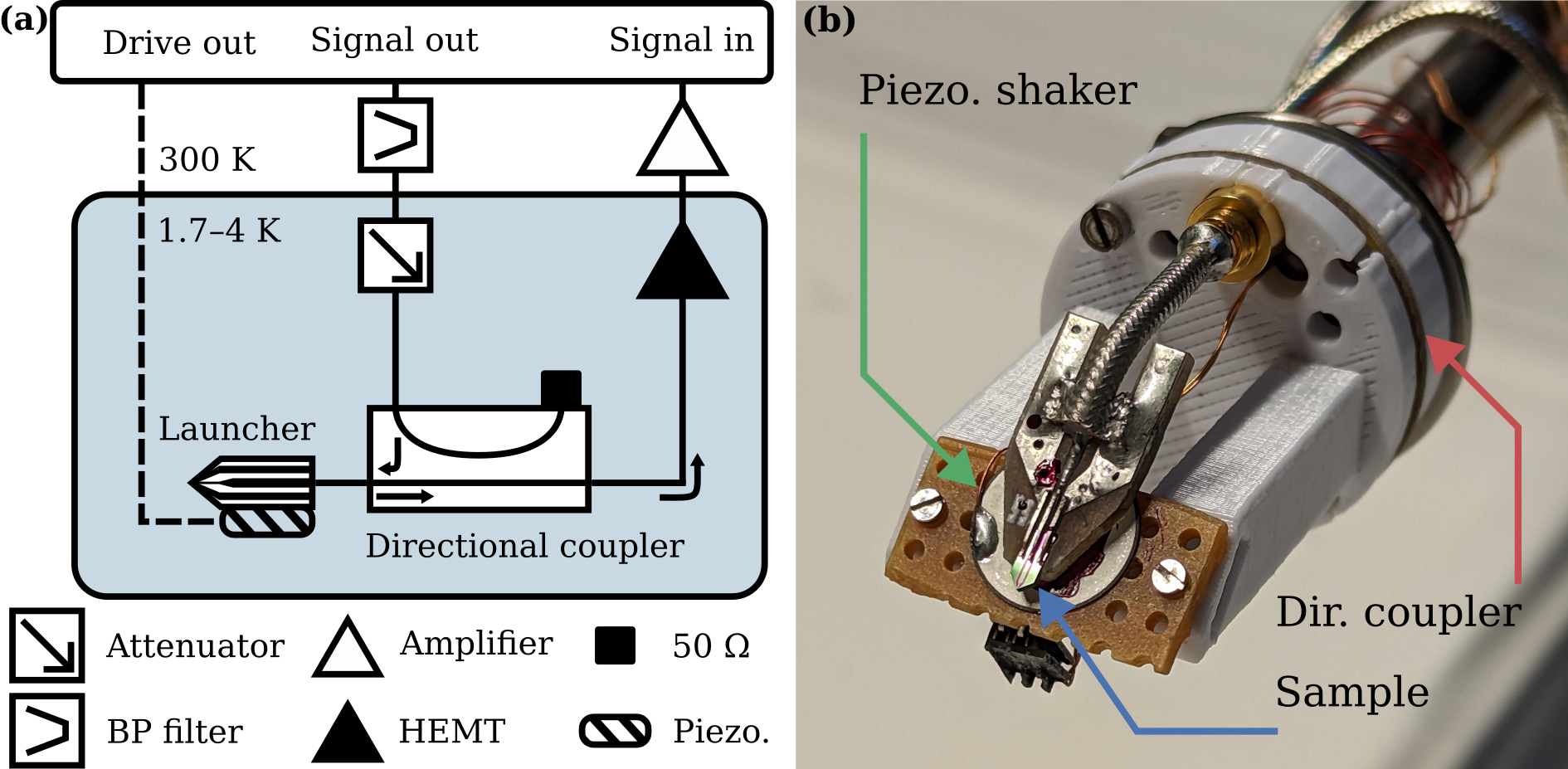}
    \caption{
        (a) Schematic of the measurement setup. The excitation signal is band-pass filtered at room temperature, attenuated at low temperature and passed to the sample. A directional coupler separates incoming and reflected signal. The reflected signal is amplified with a cryogenic low-noise amplifier and additionally amplified at room temperature. A separate port on the microwave platform generates the low-frequency drive to a piezoelectric shaker.
        (b) Photograph of the sample, home-made launcher and directional coupler integrated at the bottom of the microwave inset. The piezoelectric shaker used to intertially actuate the cantilever of the sample (held by three prongs at the front end of the launcher) is visible as the white disk beneath.
    }
    \label{fig:measurement_setup}
\end{figure*}

\subsection{Electrical mode}\label{sec:electrical_characterization}

From the measured normal-state resistance of our nanowires and the measured thickness and width, we find a sheet resistance $R_{\square}$ = \SI{243}{\ohm\per\sq}, corresponding to a resistivity of $\rho_n$ = \SI{365}{\micro\ohm\centi\meter}. We monitor the microwave response during cool-down and estimate a critical temperature $T_c$ = \SI{9.6}{\kelvin}, from which we estimate the superconducting energy gap with the BCS relation $\Delta_{0} = 1.76 \kB T_c$ = \SI{1.46}{\milli\electronvolt}. Using Eqn.~\eqref{eqn:kinetic_per_square} we find a kinetic inductance per square $\Lks$ = \SI{35}{\pico\henry\per\sq} for the 200-nm-wide nanowires, corresponding to a kinetic inductance per unit length $L_{k} / \ell$ = \SI{175}{\pico\henry\per\micro\meter}. We compare this to the estimated geometric inductance per unit length, using the thin-ribbon formula~\cite{Valenti2019, Amin2022} $L_{g} \approx (\mu_0/2\pi) \ell \ln (2 \ell / w)$, from which we obtain $L_{g} / \ell$ = \SI{17}{\pico\henry\per\micro\meter} for our 200-nm-wide nanowires. The ratio of kinetic inductance to total inductance $\alpha = L_{k} / (L_{k} + L_{g}) \simeq 1$, meaning that we can safely neglect the geometric contribution to the total inductance of our nanowires. This approximation is also valid for other samples with smaller nanowire widths which are expected to have a higher $L_{k} / \ell$.

We measured the microwave cavity resonant frequency $\omega_{c}$ on 26 chips. In some cases we studied the temperature dependence of $\omega_{c}$ and verified electromechanical coupling between the cavity mode and cantilever mode. These measurements were performed in a dry cryostat (DynaCool Physical Properties Measurement System from Quantum Design), with a base temperature of \SI{1.7}{\kelvin}. We modified a measurement stick by adding high-frequency coaxial cabling for microwave signals to probe the cavity response and twisted pairs for lower-frequency signals, such as the voltage applied to the piezo disk that inertially actuates the cantilever. The stick is equipped with a cold attenuator, directional coupler and cryogenic amplifier for low-noise microwave reflection measurement, as shown in Fig.~\ref{fig:measurement_setup}. Low- and high-frequency signals are synchronously synthesized and measured with a digital multifrequency microwave measurement device (Vivace from Intermodulation Products AB) to measure phase-sensitive electromechanical transduction~\cite{Roos2023kimec}.

\begin{figure*}
    \centering
    \includegraphics[width=0.9\linewidth]{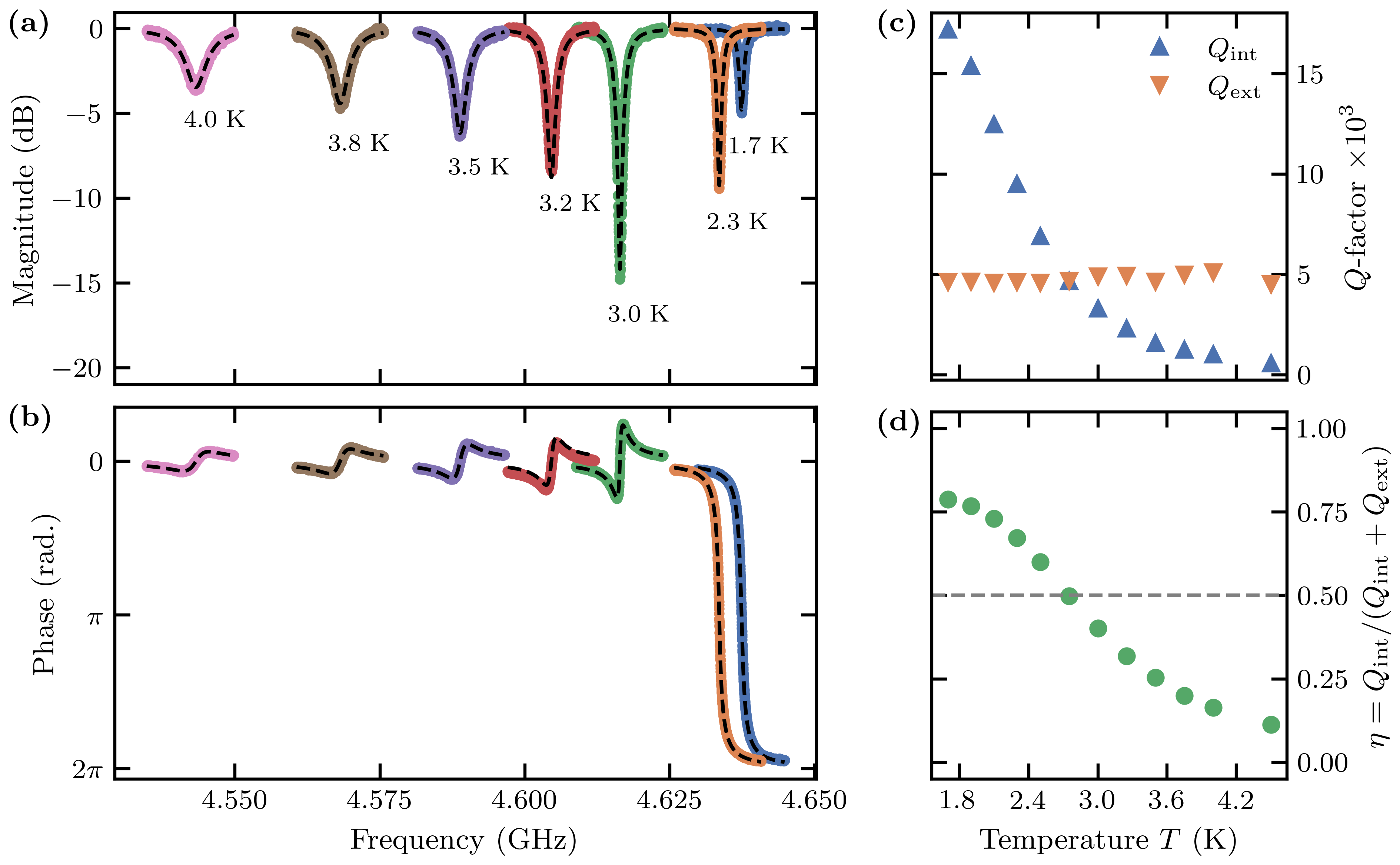}
    \caption{
        (a) The temperature-dependent magnitude and (b) phase response of the microwave resonator as a function of probe frequency with on-chip power \SI{-127}{dBm}, and with a fit to theory (dashed lines). The cavity is over-coupled at low temperatures $T \approx$~\SI{1.7}{\kelvin}, apparent from the $2\pi$-phase flip in the reflection measurement. As the temperature increases over the range \SIrange{1.7}{5}{\kelvin}, the cavity becomes critically coupled at $T$ = \SI{2.8}{\kelvin}, above which it becomes under-coupled.
        (c) The extracted internal and external quality factors $\Qint$ and $\Qext$ as a function of temperature.
        (d) The coupling parameter $\eta$ as a function of temperature.
    }
    \label{fig:temperature_sweep}
\end{figure*}

We sweep the microwave frequency and measure the reflected signal (amplitude and phase) to locate the resonance. A sharp dip in reflection is easily observed in the more slowly varying background. We zoom in on the dip to capture the resonance lineshape at low power, i.e. in the linear response regime. Using standard methods~\cite{Probst2015circlefit} we analyze the frequency dependence of the reflected amplitude and phase to determine the cavity resonant frequency $\omega_{c}$, internal quality factor $\Qint$ and external quality factor $\Qext$. From the measured $\omega_{c}$ and the simulated capacitance, we extract a kinetic inductance per square $\Lks$ for the meandering structures of different nanowire widths. For the sample shown in Fig.~\ref{fig:temperature_sweep}, we find $\omega_{c} / 2 \pi$ = \SI{4.637}{\giga\hertz}, $\Qext$ = \num{4747} and $\Qint$ = \num{17690} at $T$ = \SI{1.7}{\kelvin} with on-chip power \SI{-127}{dBm}. However, most of the tested samples were strongly over-coupled, making it difficult to extract a reliable value for $\Qint$ using the standard fitting methods. Nevertheless, we can reliably determine the resonant frequency for all measured samples and calculate the kinetic inductance using the simulated capacitance and nominal number of squares in the meander. Table~\ref{tab:parameters_per_nanowire_width} summarizes the kinetic inductance per square thus determined for different nanowire widths $w$. Values of $\Lks$, in the range \SIrange{32}{60}{\pico\henry\per\sq} across all nanowire widths, are in approximate agreement with nanowires of similar materials and dimensions~\cite{Hazard2019fluxonium, Niepce2019kinetic, Amin2022, Giachero2023films}.

\begin{table}[]
    \caption{\label{tab:parameters_per_nanowire_width}
        Summary of the mean kinetic inductance per square $\bar{L}_{k, \square}$, mean resonant frequency $\bar{\omega}_c$, and the range of measured values, grouped by the nominal nanowire width $w$.
    }
    \begin{tabular}{@{}lllllll@{}}
    \toprule
    $w$& Samples & $\bar{L}_{k, \square}$ & $\Lks$ & $\bar{\omega}_c / 2 \pi$ & $\omega_{c} / 2 \pi$ \\
    (\si{\nano\meter}) & - & (\si{\pico\henry\per\sq}) & (\si{\pico\henry\per\sq}) & (\si{\giga\hertz}) & (\si{\giga\hertz}) \\ \midrule
     75 & 8 & 52 & \numrange{41}{60} & 4.82 & \numrange{4.49}{5.34} \\
     100 & 7 & 51 & \numrange{46}{54} & 4.43 & \numrange{4.25}{4.64} \\
     200 & 11 & 34 & \numrange{32}{38} & 4.42 & \numrange{4.22}{4.56} \\ \bottomrule
    \end{tabular}
\end{table}

We also studied the temperature dependence of the microwave response in the range \SIrange{1.7}{5}{\kelvin} where we find a shift of the resonant frequency $\omega_{c}$ by many linewidths, and a change of coupling parameter $\eta$. Figures~\ref{fig:temperature_sweep}(a) and \ref{fig:temperature_sweep}(b) show the amplitude and phase of the reflected signal for one of the chips, as a function of temperature. At each temperature we fit to extract $\omega_{c}$, $\Qext$ and $\Qint$. As shown in Fig.~\ref{fig:temperature_sweep}(c), the external losses are roughly independent of temperature, while the internal quality factor degrades with temperature. The change in $\omega_{c}$ results from a temperature dependence of $L_{k}$, which, together with the change in $\Qint$, results in a transition from over-coupled to under-coupled at $T \approx$~\SI{2.8}{K}, as shown in Fig.~\ref{fig:temperature_sweep}(d).

\subsection{Electromechanical coupling}\label{sec:electromechanical_characterization}

\begin{figure}
    \centering
    \includegraphics[width=0.9\linewidth]{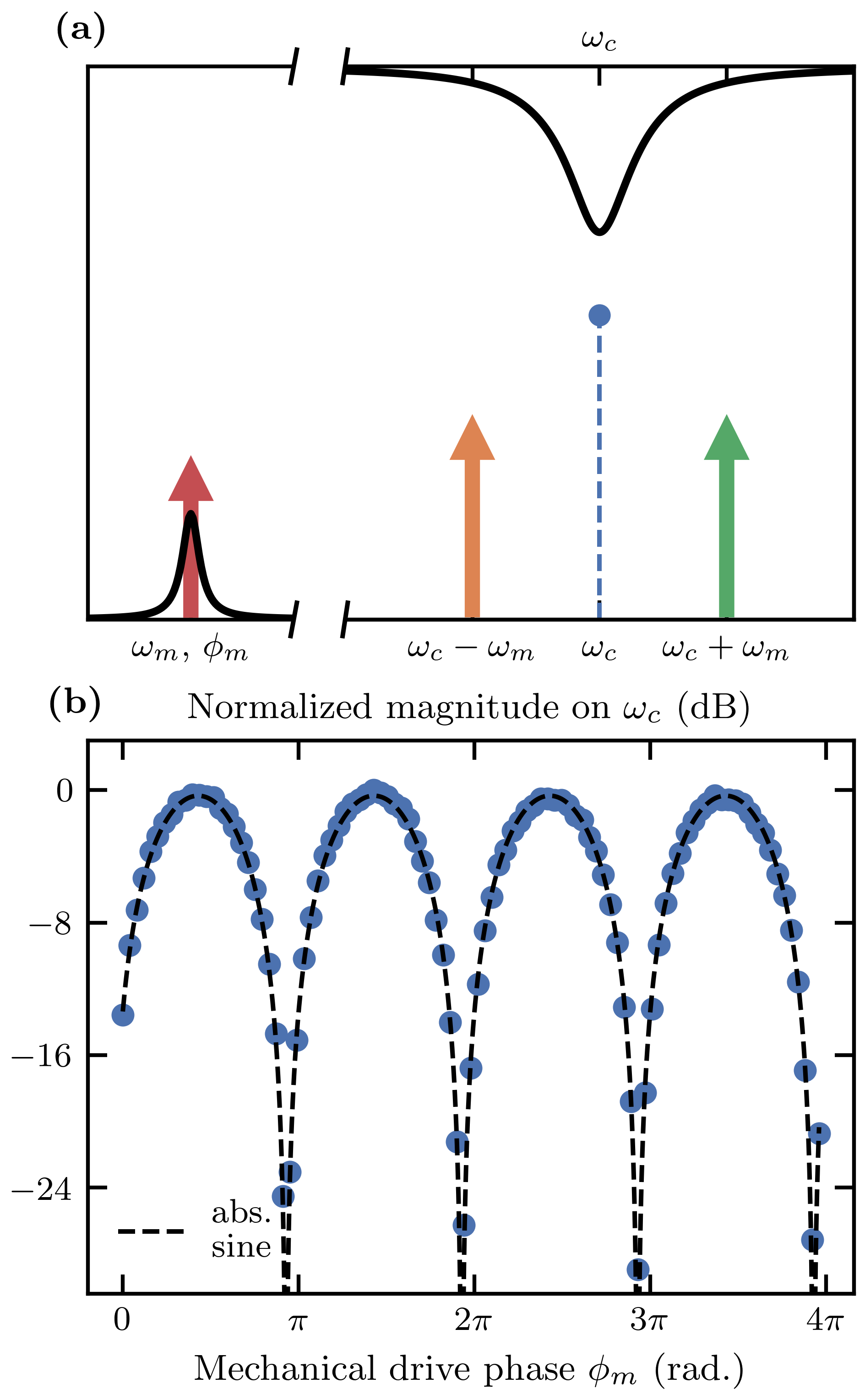}
    \caption{
        (a) Illustration of the pump-drive scheme for phase-sensitive detection of the mechanical oscillation. Two microwave tones of equal power and fixed relative phase are applied symmetrically about the cavity's resonant frequency $\omega_{c}$ and detuned by $\pm \omega_{m}$. Simultaneously, a separate tone drives the mechanical oscillator coherently at its resonant frequency $\omega_{m}$ with a variable mechanical phase $\phi_m$. The electromechanical coupling mixes the mechanical frequency with each microwave tone, leading to an interfering response at $\omega_{c}$.
        (b) Normalized measured magnitude of the response at $\omega_{c}$ as a function of the applied mechanical drive phase $\phi_m$. The interference fringes show an excellent fit to $\vert \sin \phi_m \vert$, characteristic of a balanced interferometer. 
    }
    \label{fig:electromechanical_coupling}
\end{figure}

Electromechanical coupling allows us to detect forces on the tip by measuring the bending of the cantilever. Measurements of the harmonic motion of a nanomechanical resonator coupled to a microwave electrical mode using multiple microwave tones have demonstrated very high force sensitivities~\cite{Hertzberg2010bae}. We use a multifrequency drive and read-out scheme, facilitated by device being in the sideband-resolved regime~\cite{Roos2023kimec}. Figure~\ref{fig:electromechanical_coupling}(a) illustrates the working principle, where two microwave tones are applied at $\omega_{c} \pm \omega_{m}$, while simultaneously actuating the cantilever through a coherent drive. Sidebands generated by the mechanical motion interfere with each other at the cavity's resonant frequency $\omega_{c}$, either constructively or destructively, depending on the phase of the mechanical motion $\phi_{m}$, which is a controllable parameter. Figure~\ref{fig:electromechanical_coupling}(b) shows this phase-sensitive detection of the signal as a function of the mechanical phase $\phi_m$ relative to the microwave drive tones. This measurement was made in a dilution refrigerator on a sample with $\omega_{m} / 2 \pi$ = \SI{5.828}{\mega\hertz}.

\section{Conclusion}\label{sec:conclusion}

We described our approach to designing cantilever force sensors with integrated microwave cavity electromechanical sensing of flexural motion, based on the strain-dependent kinetic inductance of a superconducting nanowire. This type of force sensor is potentially interesting for low-temperature AFM as the superconducting cavity read-out scheme is intrinsically low noise. Compared to other low-temperature AFM probes, such as the qPlus sensor~\cite{Giessibl2019qplus}, our versatile design allows for achieving a wide variety of the parameters of the mechanical mode (force transducer) with a much lower effective mass, and consequently an improved force sensitivity. The integrated resonator used for motion detection also provides a potentially quantum-limited approach to motion detection. It is difficult to make a complete and detailed comparison of AFM force sensors using a microwave resonator, with those using an optical cavity to detect motion~\cite{Srinivasan2011cantilever, Schwab2022afm} due to numerous differences in the entire measurement apparatus. However, in comparison to optical resonators, the ability of microwave resonators to easily operate in the sideband-resolved regime allows for additional and potentially more efficient schemes of sensing force with minimal back-action from motion detection. 

Our design covers a vast parameter space, balancing different considerations for both the electrical mode, such as the critical current $I_{c}$, critical temperature $T_{c}$ and kinetic inductance per square $\Lks$; and the mechanical mode, such as the resonance frequency $\omega_{m}$, quality factor $Q_{m}$ and spring constant $k$ of the cantilever. These trade-offs affect the transduction efficiency and force sensitivity. Although a large single-photon coupling rate $g_{0}$ is desirable, we prioritize a cavity that we can pump to large intra-cavity photon number while maintaining linearity. In this regard, variations on the design presented here need to be fabricated and tested. The sensors described here represent the first generation of devices, where there is room for improvement. These devices served to establish the fabrication process that we described in detail herein, and to verify that kinetic inductive mechano-electric coupling (KIMEC) as a useful, albeit not entirely understood, physical effect. Further investigation of alternative designs will help to shed light on the underlying physical mechanism behind KIMEC. We hope that this study motivates future work in this direction.

\section*{Acknowledgments}
We thank the Quantum-Limited Atomic Force Microscopy (QAFM) team for fruitful discussions: T Glatzel, M Zutter, E Tholén, D Forchheimer, I Ignat, M Kwon, and D Platz. The European Union Horizon 2020 Future and Emerging Technologies (FET) Grant Agreement No. 828966 --- QAFM and the Swedish SSF Grant No. ITM17--0343 supported this work.

\section*{Data availability}
The data generated and analyzed during this study is openly available in Zenodo at \url{https://doi.org/10.5281/zenodo.8246258}.

\section*{Competing interests}
D.B.H. is a part owner of Intermodulation Products AB, which manufactures and sells the microwave measurement platform used in this experiment. A.K.R, E.S., E.K.A., E.H. declare no competing interests.

\bibliographystyle{unsrtnat}

\end{document}